\documentclass[epj]{svjour}
\usepackage{graphics}
\newcommand{\msb}{\overline{\mathrm{MS}}}
\newcommand{\Lr}{\Lambda}
\newcommand{\slap}[1]{\rlap{$#1$}\hspace{0.3ex}/}

\newcommand{\sla}[1]{\rlap{$#1$}/}
\newcommand{\be}{\begin{equation}}
\newcommand{\bea}{\begin{eqnarray}}
\newcommand{\ee}{\end{equation}}
\newcommand{\eea}{\end{eqnarray}}

\begin{document}
\title{Power corrections in models with extra dimensions}
\author{J.F. Oliver\and J. Papavassiliou  \and A. Santamaria 
}                     
%
%
\institute{ Departamento de F\'{\i}sica Te\'orica and IFIC, 
Universidad de Valencia-CSIC, 
E-46100, Burjassot, Valencia, Spain }
\date{\today}
%
\abstract{
We critically revisit the issue of power-law running 
in models with extra dimensions. The  general
conclusion is  that, in the  absence of any additional  
physical principle, the
power-corrections tend to  depend strongly on the details
of the underlying theory.
} 

\maketitle
\section{Introduction}
The power-law running of couplings has been considered 
as one of the most characteristic predictions of models with extra dimensions
\cite{Arkani-Hamed:1998nn},
allowing the exciting possibility of an early unification
\cite{Dienes:1998vg}. Even though there is no doubt that 
power-law corrections will appear in such theories, their precise 
physical interpretation merits further scrutiny 
\cite{Contino:2001si,Oliver:2003cy}. 

The basic argument in favor of power-law running is 
\( \msb  \) inspired. 
In a general renormalization scheme satisfying decoupling 
the $\beta$ function assumes the form 
\begin{equation}
\beta =\sum _{n} \beta _{0}f \bigg(\frac{\mu }{M_{n}}\bigg)
\label{beta}
\end{equation}
with $\mu$ the renormalization scale, 
$\beta_{0}$ the contribution
of a single mode, and
\( f(\mu /M)\rightarrow 0\, \, \, \, \, \mu \ll M \)
and \( f(\mu /M)\rightarrow 1\, \, \, \, \, \mu \gg M \).  
In particular, in the case of the \( \msb  \)
the function $f(\mu /M)$ is chosen to be the step-function,  
\( f(\mu /M)\equiv \theta (\mu /M-1) \), in order to enforce decoupling. 
Theories with \( \delta  \) extra compact dimensions contain
an infinite tower of Kaluza-Klein (KK) modes with masses 
\begin{equation}
 M^{2}_{n}=\left( n^{2}_{1}+n^{2}_{2}+\cdots +n^{2}_{\delta }\right) M^{2}_{c} ,
\label{KKmasses}
\end{equation}
where \( M_{c}=1/R_{c} \) is the compactification scale.
Then, the 
naive way 
of generalizing the \( \msb  \) in the presence of 
an infinite number of such modes
is simply
\begin{eqnarray*}
\beta =\sum _{n<\mu /M_{c}}\beta _{0}\approx \beta _{0}\int _{n<\mu /M_{c}}d\Omega _{\delta }n^{\delta -1}dn &  & \\
=\beta _{0}\frac{1}{\Gamma (1+\delta /2)}
\left( \pi \frac{\mu ^{2}}{M^{2}_{c}}\right) ^{\delta /2} , &  & 
\end{eqnarray*}
giving rise to a $\beta$ which 
just counts the number of active modes, 
i.e. lighter than $\mu$.
But this generalization is ambiguous, because the \( \msb  \) 
scheme does not satisfy decoupling. Instead, decoupling
must be imposed by hand every time a threshold is crossed \cite{Georgi:qn}. 
Therefore,
in the presence of an infinite number of thresholds
the result becomes extremely dependent on the prescription used.

Of course, particles decouple naturally and smoothly 
in the Vacuum Polarization Function (VPF), because of unitarity
(optical theorem). In \cite{Dienes:1998vg}
the VPF 
of the photon was calculated 
in the presence of the tower of fermionic KK modes. However,  
the VPF was computed at \( Q^{2}=0 \), a fact which   
obscures the relation with the optical theorem.  
In addition, a cutoff \( \Lambda  \) (in proper
time) was used.
The cutoff was eventually identified with the 
sliding scale and
the result used to compute the $\beta$ function. It is easy to convince oneself 
however, that the above  
procedure is equivalent to using the function 
\( f(\Lambda /M)\equiv \exp \left( -M^{2}_{n}/\Lambda ^{2}\right)  \)
to decouple the KK modes. So, the resulting $\beta$-function  reads
\[
\beta =\sum _{n}\beta _{0}e^{-\frac{M^{2}_{n}}{\Lambda ^{2}}}\approx \beta _{0}\left( \pi \frac{\Lambda ^{2}}{M^{2}_{c}}\right) ^{\delta /2} ,\] 
and $\mu$ was chosen by hand to satisfy
\( \mu ^{\delta }=\Gamma (1+\delta /2)\Lambda ^{\delta } \)
in order to reproduce the \( \msb  \) inspired result.\\ 
\noindent
Thus, even though the VPF is used, the introduction of 
a hard cutoff is not any better conceptually
than the direct use of a sharp step function for decoupling the modes:
one gets a smooth $\beta$ function because one puts in by hand 
a smooth function to decouple the KK modes. 
Because of the very sharp step-like decoupling, these two
ways of decoupling KK modes lead to a finite result for any
number of extra dimensions. 

The physical decoupling
function \( f(\mu /M) \)
that is really obtained from the VPF 
can be approximated \cite{Brodsky:1998mf} by 
the simple expression
\( f(\mu /M)=\mu ^{2}/(\mu ^{2}+5M^{2}) \). Substituting it in 
Eq.(\ref{beta}), we see that 
the sum over all KK modes converges only for 
one extra
dimension, but is badly divergent for several extra
dimensions. 
The extra infinities one finds when summing all the KK modes are just the manifestation
of the non-renormalizability of the underlying uncompactified
theory; the latter is non-renormalizable simply because 
the gauge coupling has dimension
\( 1/M^{\delta /2} \). 
Therefore, 
higher dimension operators are needed as counterterms, and one is naturally led 
to the effective field theories (EFT) .

\section{Extra dimensions and  EFT}
The general rules 
of continuum EFT (to be distinguished from the 
``Wilsonian'' type \cite{Wilson:1973jj}), may be summarized as follows \cite{Georgi:qn}: 
(i) Virtual momenta in loops run 
up to infinity; (ii) heavy particles are 
removed from the spectrum at low energies; 
(iii) effects of heavy particles are absorbed in the coefficients 
of higher-dimensional operators;
(iv)
regularization and renormalization are necessary; 
(v) the use  of dimensional regularization
and the \( \msb  \) scheme is advantageous, because 
in that case there is  
no mixing between operators of different dimensionality.

We next proceed to use the continuum EFT at the level of 
the extra-dimensional (uncompactified) theory. 
Clearly, the virtual 
momenta associated with the extra dimensions run up to infinity,
as in point (i) above. 
However, at the level of the 4-d (compactified) theory they are 
KK masses
and one is supposed to keep only particles lighter 
than the relevant scale. 
Thus, truncating the KK tower 
by introducing a large (but otherwise arbitrary) cutoff $N_s$
amounts to cutting-off the momenta of the 
uncompactified theory. Identifying $N_s$ 
with a physical cutoff gives  illusion
of predictivity, but is plagued with ambiguities \cite{Burgess:1992gx}. 
Even when using cutoffs one has to add counterterms from higher
order operators, absorb the cutoff and express the result
in terms of a series of unknown coefficients.
Therefore, in order to define a 
genuine non-cutoff {continuum} EFT framework, 
we must keep {\it all} KK modes, or, 
equivalently, study how they decouple {\it all of them at once}.

\section{Computing the VPF in a toy model}
We will follow the strategy outlined above in the context of a simple toy model 
\cite{Oliver:2003cy}.
Consider a theory with one fermion and 
one photon in \( 4+\delta  \)
dimensions, with the extra dimensions compactified
on a torus of equal radii \( R_{c}\equiv 1/M_{c} \).
The Lagrangian reads  
\be
{\mathcal{L}}_{\delta }=-\frac{1}{4}F^{MN}F_{MN}+i\bar{\psi }\gamma ^{M}D_{M}\psi +\mathcal{L}_{\mathrm{ct}}
\ee
where \( M=0,\cdots ,3,\cdots ,3+\delta  \),  
\( \mu =0,\cdots \, 3 \), 
\( D_{M}=\partial _{M}-ie_{D}A_{M} \), and \( e_{D} \) is the coupling in 
\( D=4+\delta  \)
dimensions. It has dimension \( [e_{D}]=1/M^{\delta /2} \). 
After compactification, the four-dimensional dimensionless 
gauge coupling, \( e_{4} \), and \( e_{D} \)
are related by the compactification scale: 
\( e_{4}=e_{D}\left( \frac{M_{c}}{2\pi }\right) ^{\delta /2} \).
The part of the spectrum relevant to our purposes is:
(i) one massless photon; (ii)
\( 2^{[\delta /2]} \) massless Dirac fermions; (iii)
a tower of massive Dirac fermions with masses
given by Eq.(\ref{KKmasses}). In addition, 
the counterterm Lagrangian is given by
$
\mathcal{L}_{\mathrm{ct}}=\frac{\kappa _{1}}{M_{s}}D_{M}F^{MK}D^{N}F_{NK}+\cdots$, 
where the ellipses denote operators of higher dimensionality.

We next compute the VPF at the level of 
the compactified theory; at one-loop it is given by 
\[
\Pi ^{\mu \nu }(q)=ie^{2}_{4}\sum _{n}\int \frac{d^{4}k}{(2\pi )^{4}}
\mathrm{Tr}\left\{ \gamma ^{\mu }\frac{1}{\sla k-M_{n}}\gamma ^{\nu }\frac{1}{\sla k+\sla q-M_{n}}\right\} .\]
From gauge invariance, $\Pi ^{\mu \nu }(q) = 
\left( q^{2}g^{\mu \nu}-q^{\mu}q^{\nu }\right) \Pi(q)$.
In order to exploit dimensional regularization techniques, we   
add and subtract the contribution of VPF
in the uncompactified space: $\Pi (q) = 
\left[\Pi (q)-\Pi _{\mathrm{uc}}(q)\right] + \Pi _{\mathrm{uc}}(q)$ .
This allows us to trade off the divergent sum for a divergent integral,
which may be computed using the standard results of dimensional regularization.
Indeed, 
one may verify that $\left[\Pi (q)-\Pi _{\mathrm{uc}}(q)\right]$
is in fact UV and IR finite (and can be evaluated numerically), 
whereas the expression 
\begin{equation}
\Pi^{MN}_{\mathrm{uc}}(q)=ie^{2}_{D}\int \frac{d^{4+\delta }k}{(2\pi )^{4+\delta }}
\mathrm{Tr}
\left\{ \gamma ^{M}\frac{1}{\slap k}\gamma ^{N}\frac{1}{\slap k+\slap q}\right\} 
\nonumber
\label{PUC}
\end{equation}
contains all divergent contributions. In particular, setting 
$\Pi ^{MN}_{\mathrm{uc}}(q)=\left( q^{2}g^{MN}-q^{M}q^{N}\right) \Pi _{\mathrm{uc}}(q)$, 
we have that 
\begin{equation}
\Pi _{\mathrm{uc}}(Q)=
\frac{e^{2}_{4}}{2\pi ^{2}}\frac{\pi ^{\delta /2}\, 
\Gamma ^{2}(2+\frac{\delta }{2})}{\Gamma (4+\delta )}
\Gamma \left( -\frac{\delta }{2}\right) \, 
\left( \frac{Q^{2}}{M_{c}^{2}}\right) ^{\delta /2} .
\label{dimreg}
\end{equation}
The extra divergences that cannot be canceled by the wave-function 
renormalization are to be  
absorbed in the operator
\( D_{M}F^{MK}D^{N}F_{NK} \). 
In the limit \( Q^{2}\ll M^{2}_{c} \), we finally obtain \cite{Oliver:2003cy}
\[
\Pi ^{(\delta )}(Q)=\frac{e_{4}^{2}}{2\pi ^{2}}
\bigg( a^{(\delta )}_{0}-
\underbrace{\frac{1}{6}
\log \left(\frac{Q^{2}}{M^{2}_{c}}\right)}_{{ordinary ~running}} 
+a^{(\delta )}_{1}\frac{Q^{2}}{M^{2}_{c}}+\cdots \bigg) 
\]
with
\vspace{0.3cm}
{\centering \begin{tabular}{|c|c|c|c|}
\hline 
\( \delta  \)&
\( 1 \)&
\( 2 \)&
\( 3 \)\\
\hline 
\hline 
\( a^{(\delta )}_{0} \)&
\( -0.335 \)&
\( -0.159 \)&
\( -0.094 \)\\
\hline 
\( a^{(\delta )}_{1} \)&
\( -0.110 \)&
\( 0.183 \)&
\( 0.298 \)\\
\hline 
\end{tabular}\par}
\vspace{0.3cm}
Notice that the 
coefficients \( a^{(\delta )}_{1} \) can be
affected by non-calculable contributions from higher dimension
operators. 

Using the above VPF we define a sort of ``effective charge''
\begin{equation}
\frac{1}{\alpha _{\mathrm{eff}}(Q)}\equiv \frac{1}{\alpha _{4}}\left( 1+\Pi ^{(\delta )}(Q)\right) . 
\end{equation}
To determine the relation between \( e_{4} \) and the QED
coupling, we identify our effective charge at some low energy scale (for instance
\( Q^{2}=m^{2}_{Z}\ll M^{2}_{c} \))
with the QED coupling
\begin{equation}
\frac{1}{\alpha _{\mathrm{eff}}(m_{Z})}\approx \frac{1}{\alpha _{4}}+\frac{2}{\pi }a^{(\delta )}_{0}-\frac{2}{3\pi }\log \left( \frac{m_{Z}}{M_{c}}\right) . 
\end{equation}
The relation between the QED coupling 
\( \alpha _{\mathrm{eff}}(m_{Z}) \)
and \( \alpha _{4} \) 
contains only logarithmic running. 
This is the only matching we
can reliably compute without knowing the physics beyond 
\( M_{s} \). 
Notice also that, in this EFT framework, the gauge coupling does not run 
above the compactification scale.
That is
what happens in \( \chi PT \) \cite{Pich:1995bw}, 
when using dimensional regularization :
\( f_{\pi } \) does not run, 
it just renormalizes higher dimensional operators. 

For all energies we can write for the effective coupling:
\[
\frac{1}{\alpha _{\mathrm{eff}}(Q)}
\approx \frac{1}{\alpha _{\mathrm{eff}}(m_{Z})}
+\frac{1}{\alpha _{4}}\left( \Pi ^{(\delta )}(Q)-\Pi ^{(\delta )}(m_{Z})\right). \]

\begin{figure}
\resizebox{0.48\textwidth}{!}{%
\includegraphics{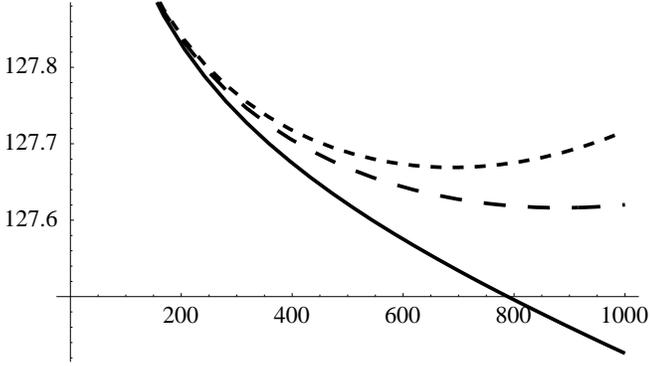}} 
\caption{The effective coupling as a function of $Q$, for 
$\delta=1$ (solid), $\delta=2$ (long-dashed), and $\delta=3$ (short-dashed)}
\label{fig:1}
\end{figure}

As seen in Fig.\ref{fig:1}, at low energies it displays the standard 
logarithmic running, whereas 
at \( Q^{2}/M^{2}_{c}\approx 1 \)
the behavior deviates dramatically from the logarithmic
running. 
However, for \( Q>M_{c} \) this effective
charge cannot 
be interpreted anymore as the running charge
since it could contain physics from higher dimension operators.

We finally comment on possible pitfalls related to the 
cavalier interpretation of hard cutoffs in terms of physical masses, especially 
if the theory that gives meaning to those masses is not known \cite{Burgess:1992gx}. 
Cutoffs can give an indication of the presence
of power corrections, but the coefficients of these corrections
cannot be computed without knowing the details
of the full theory. Results obtained through
the use of hard cutoffs hint to the appearance of contributions
which go as \( \left( M_{s}/M_{c}\right) ^{\delta } \), where $M_{s}$
is some scale related to the onset of new physics;
however, 
the coefficients multiplying these corrections are not reliably determined.

To illustrate this point with an example, let us 
calculate the expression of Eq.(\ref{PUC}) using a cutoff $\Lr$:
\begin{eqnarray}
\Pi ^{(1)}_{\mathrm{uc}}(Q) &=&\frac{e_{4}^{2}}{2\pi ^{2}}\left( -\frac{3\pi ^{2}Q}{64M_{c}}+\frac{\sqrt{\pi }Q^{2}}{15M_{c}}+\frac{\sqrt{\pi }\Lr }{3M_{c}}\right) 
\nonumber\\
\Pi ^{(2)}_{\mathrm{uc}}(Q)&=&\frac{e_{4}^{2}}{2\pi ^{2}}\left( \frac{\pi \Lr ^{2}}{6M_{c}^{2}}+\frac{\pi Q^{2}}{30M_{c}^{2}}
\left[ \log (Q^{2}/\Lr ^{2})+\gamma -\frac{77}{30}\right] \right)
\nonumber\\
\Pi ^{(3)}_{\mathrm{uc}}(Q)&=&\frac{e_{4}^{2}}{2\pi ^{2}}\left( \frac{5\pi ^{3}Q^{3}}{768M_{c}^{3}}-\frac{\pi ^{3/2}Q^{2}\Lr }{15M_{c}^{3}}+
\frac{\pi ^{3/2}\Lr ^{3}}{9M_{c}^{3}}\right) .
\label{hard}
\end{eqnarray}
We note that 
the terms independent of the cutoff are the same as those obtained by using
dimensional regularization, Eq.(\ref{dimreg}),  
 whereas the 
additional pieces depend strongly on the cutoff. 
Let us next assume that the ``new physics'' 
is due to the presence of an additional fermion 
in our \( 4+\delta  \) dimensional
theory, whose mass satisfies
\( M_{f}\gg M_{c} \), and let us compute its effects on the coupling
constant for \( M_{c}\ll Q^{2}\ll M_{f} \): We have
\begin{eqnarray}
\Pi ^{(\delta )}_{f}(Q) &=&\frac{e_{4}^{2}}{2\pi ^{2}}\left( \frac{\pi }{M_{c}}\right) ^{\delta /2}\Gamma (-\delta /2)
\nonumber\\
&\times& \int dxx(1-x)\left( M^{2}_{f}+x(1-x)Q^{2}\right) ^{\delta /2}
\end{eqnarray}
Expanding for \( Q^{2}\ll M_{f} \) and integrating on \( x \) 
we obtain
\begin{equation}
\Pi ^{(\delta )}_{f}(Q)\approx \frac{e_{4}^{2}}{2\pi ^{2}}\left( \sqrt{\pi }\frac{M_{f}}{M_{c}}\right)^{\delta }\Gamma \left( -\frac{\delta }{2}\right) \left( \frac{1}{6}+\frac{\delta }{60}\frac{Q^{2}}{M_{f}^{2}}\right) .
\end{equation}
Evidently, when integrating out the heavy fermion one receives 
power corrections
to the gauge coupling ( also higher dimension operators ), 
even if the dimensional regularization is employed.
However, the coefficients are completely different from those 
obtained in Eq.(\ref{hard}) using a hard cutoff.

\section{Conclusions}

We have argued that 
continuum EFT, with dimensional regularization and the \( \msb  \) scheme, 
provides a self-consistent 
framework for computing in models with extra dimensions. 
The running of the coupling that can be 
computed reliably within this framework 
is only logarithmic. 
Additional power corrections are expected, but
cannot be computed without knowing the details
of the complete theory in which the \( D \) dimensional theory is  
embedded (for example, extra-dimensional GUT).
Thus, the requirement of coupling unification opens a window to
physics much beyond the compactification
scale.

\medskip

{\it Acknowledgments}:
This work has been supported by the Spanish MCyT under the grants 
BFM2002-00568 and FPA2002-00612, and by the OCyT  
of the {}``Generalitat Valenciana{}'' under the Grant GV01-94.


\begin{thebibliography}{99}
\bibitem{Arkani-Hamed:1998nn}
N.~Arkani-Hamed, S.~Dimopoulos and G.~R.~Dvali,
Phys.\ Rev.\ D {\bf 59}, 086004 (1999); 
Phys.\ Lett.\ B {\bf 429}, 263 (1998);
I.~Antoniadis,
Phys.\ Lett.\ B {\bf 246}, 377 (1990);
C.~P.~Bachas,
JHEP {\bf 9811}, 023 (1998).

\bibitem{Dienes:1998vg}
K.~R.~Dienes, E.~Dudas and T.~Gherghetta,
Nucl.\ Phys.\ B {\bf 537}, 47 (1999).

\bibitem{Contino:2001si}
R.~Contino, L.~Pilo, R.~Rattazzi and E.~Trincherini,
Nucl.\ Phys.\ B {\bf 622}, 227 (2002).

\bibitem{Oliver:2003cy}
J.~F.~Oliver, J.~Papavassiliou and A.~Santamaria,
Phys.\ Rev.\ D {\bf 67}, 125004 (2003).

\bibitem{Georgi:qn}
H.~Georgi,
Ann.\ Rev.\ Nucl.\ Part.\ Sci.\  {\bf 43}, 209 (1993);
A.~V.~Manohar,
arXiv:hep-ph/9606222.


\bibitem{Brodsky:1998mf}
S.~J.~Brodsky, M.~S.~Gill, M.~Melles and J.~Rathsman,
Phys.\ Rev.\ D {\bf 58}, 116006 (1998).


\bibitem{Wilson:1973jj}
K.~G.~Wilson and J.~B.~Kogut,
Phys.\ Rept.\  {\bf 12}, 75 (1974).

\bibitem{Burgess:1992gx}
C.~P.~Burgess and D.~London,
Phys.\ Rev.\ D {\bf 48}, 4337 (1993).

\bibitem{Pich:1995bw}
A.~Pich,
Rept.\ Prog.\ Phys.\  {\bf 58}, 563 (1995).


\end{thebibliography}
\end{document}